\documentclass[pre,twocolumn,notitlepage,superscriptaddress,color,showpacs]{revtex4-2}
\usepackage{amsmath,amsfonts,amssymb}
\usepackage[pdftex]{graphicx}
\usepackage{xcolor,comment}

\newcommand{\be}{\begin{equation}}
\newcommand{\ee}{\end{equation}}
\newcommand{\bd}{\begin{displaymath}}
\newcommand{\ed}{\end{displaymath}}
\newcommand{\BE}{\begin{eqnarray}}
\newcommand{\EE}{\end{eqnarray}}

\begin{document}
\title{Analysis of a voter model with an evolving number of opinion states}
\author{Jeehye Choi}
\affiliation{Advanced-Basic-Convergence Research Institute, Chungbuk National University, Cheongju, Chungbuk 28644, Korea}
\author{Byungjoon Min}
\email{bmin@cbnu.ac.kr}
\affiliation{Advanced-Basic-Convergence Research Institute, Chungbuk National University, Cheongju, Chungbuk 28644, Korea}
\affiliation{Department of Physics, Chungbuk National University, Cheongju, Chungbuk 28644, Korea}
\affiliation{Department of Medicine, University of Florida, Gainesville, FL 32610, USA}
\author{Tobias Galla}
\email{tobias.galla@ifisc.uib-csic.es }
\affiliation{Instituto de F\'isica Interdisciplinar y Sistemas Complejos IFISC (CSIC-UIB), 07122 Palma de Mallorca, Spain}
\date{\today}\date{\today}

\begin{abstract}
In traditional voter models, opinion dynamics are driven by interactions between individuals, where an individual adopts the opinion of a randomly chosen neighbor. However, these models often fail to capture the emergence of entirely new opinions, which can arise spontaneously in real-world scenarios. Our study introduces a novel element to the classic voter model: the concept of innovation, where individuals have a certain probability of generating new opinions independently of their neighbors' states. This innovation process allows for a more realistic representation of social dynamics, where new opinions can emerge and old ones may fade over time. Through analytical and numerical analysis, we find that the balance between innovation and extinction shapes the number of opinions in the steady state. 
Specifically, for low innovation rates, the system tends toward near-consensus, while higher innovation rates lead to greater opinion diversity. We also show that network structure influences opinion dynamics, with greater degree heterogeneity reducing the number of opinions in the system.
\end{abstract}
\maketitle

\section{Introduction}

The study of opinion dynamics is an active area in the field of statistical physics \cite{krapivsky1992kinetics,liggett1999stochastic,dornic2001critical,baxter2008fixation,krapivsky2010kinetic,redner2019reality}, with applications to the social \cite{castellano2009statistical,palermo2024spontaneous}, biological \cite{clifford1973model,blythe2007stochastic}, and political sciences \cite{fernandez2014voter,meyer2024time}.
Among many other approaches, the voter model (VM) has been used to model how opinions spread within a population, and to describe the dynamics of opinion formation
in social systems \cite{suchecki2004conservation,vazquez2008analytical,redner2019reality}. 
The VM is stylised and, due to its simplicity, analytically tractable with methods from statistical physics and the theory of stochastic processes \cite{redner2019reality}. Its analysis has revealed new universality classes \cite{dornic2001critical}, and interesting types of collective behaviors. The VM is also related to stochastic birth-death processes in evolutionary biology \cite{ewens2004mathematical}. In this analogy opinion states are the equivalent of alleles. More precisely, the VM describes a Moran-type process with neutral selection \cite{moran1958,donnelly1984}, that is none of the alleles has an intrinsic selective advantage over any other.

The conventional VM describes a population of a finite number of individuals, who interact all-to-all, on a regular structure such as a square lattice, or on a network \cite{castellano2003incomplete,vilone2004solution,suchecki2005voter,sood2005voter,castellano2005comparison,sood2008voter,pugliese2009heterogeneous}. In the most simple setup, each of these individuals holds a binary opinion. The dynamics unfolds though imitation; at each step a randomly chosen agent adopts the opinion of a randomly chosen neighbor. In the long-time limit, any finite system tends to one of two consensus states, where all agents share
the same opinion.

The original VM has been extended in many different ways. This includes variants of the VM with `noise', allowing individuals to change opinion states spontaneously \cite{carro2016noisy, peralta2018analytical, herrerias2019consensus}, nonlinear voter models \cite{castellano2009nonlinear,schweitzer2009nonlinear,ramirez2024ordering}, multistate voter models (models with more than two opinion states) \cite{starnini2012ordering,pickering2016,vazquez2019,nowak2021,berrios2021switching,iannelli2022,ramirez2022local}, the coevolving voter model \cite{holme2006nonequilibrium,vazquez2008generic,min2017fragmentation} in which agents can re-wire links in the interaction network, and higher-order voter models  \cite{kim2024}. 

In most of this work, the set of possible opinions that an agent can hold is fixed. Here, we introduce and analyse a VM in which new opinion states can emerge as the dynamics proceeds. This models the emergence of new information, external influences, and changes
in individuals' beliefs \cite{bornholdt2011,baek2013}. The appearance of new opinions enriches the dynamics, adding a new layer of complexity. There is also an interesting connection  to models of the evolution of `mating types' in biology \cite{Constable, berrios2021switching}. These are the equivalent of `sexes' for microorganisms. In contrast to organisms with true sexes however the number of mating types is not necessarily restricted to two \cite{Lehtonen}, in some funghi their number can be in the thousands. The evolution of mating types has been described with models of sexual or asexual reproduction. As we will see below, there are close mathematical relations between these models and the VM with dynamic number of opinion states that we study here.

The remainder of the paper is organized as follows. In Sec.~II we introduce the VM with evolving opinions and define the detailed dynamics. Section III focuses on the dynamics of the model on a complete graph. 
We analyze the behavior resulting from the effect of innovation rate through the number of opinions, the distributions of opinion sizes, and the density of the active interface. We find that when the innovation rate is low, the system remains in near-consensus states, but as the rate increases, opinion diversity grows.
In Sec.~IV we extend the analysis to random networks. We observe that heterogeneity in the degree distribution reduces the number of opinions in the system. Finally, in Sec. V we summarize and discuss our findings.

\section{Model}
\subsection{Setup and dynamics}\label{sec:setup}
The model describes a population of $N$ individuals, labeled $i=1,\dots,N$. We initially introduce a discrete-time version of the dynamics, writing $\tau$ for the number of microscopic time steps that have elapsed. At each time $\tau$
each individual holds one of a discrete number of opinions. These opinions are labeled by
an integer number $a=1,2,3,\dots$, and we write $\sigma_i(\tau)$ for the opinion state of individual
$i$ at time $\tau$. Although we will initially consider populations with all-to-all interaction, the
individuals can in principle sit on an interaction network. This means that each individual
can only interact with a set of nearest neighbours. In the case of all-to-all interaction any
individual is a nearest neighbour of all other individuals.

The dynamical rules of the model with evolving sets of opinions are similar to those of the
conventional voter model. The key difference is the possibility for an individual to introduce
a new opinion state into the population.

In detail the dynamics are as follows:
\begin{enumerate}
\item Initially ($\tau=0$), there are $M(\tau=0)=2$ opinions present in the population. Each individual is randomly assigned one of these states, each with probability $1/2$, and with no correlations between different individuals.
\item Assume $\tau$ microscopic time steps have elapsed. There are $M(\tau)$ opinions in the population, and the states of the agents are $\sigma_i(\tau)$. An individual is now chosen at random from the population (with equal probabilities across individuals). 
We label this individual $i_0$. A nearest neighbour of $i_0$ is selected at random (chosen with equal probabilities among all nearest neighbours of $i_0$). We call this neighbour $j$. Individual $i_0$ then adopts the state of $j$, i.e., $\sigma_{i_0}(\tau+1)=\sigma_j(\tau)$. All other individuals remain in their current state, i.e., $\sigma_\ell(\tau+1)=\sigma_\ell(\tau)$ for all $\ell\neq i_0$.
\item  With probability $\alpha$, a new opinion is introduced into the population (with probability $1-\alpha$ proceed directly to step 4). This occurs as follows: A random individual is chosen from the population. This individual conceives of a new opinion not currently present in the population, and assumes this state. 
\item Go to step 2.
\end{enumerate}
We note that opinion states can be re-labelled after each cycle of steps 2 and 3, such that all $\sigma_i(\tau)\in\{1,\dots,M(\tau)\}$ at all times.

We note that the dynamics in step 2  means that an opinion may go extinct from the population (if individual $i_0$ is the last individual to hold a particular opinion). Conversely, step 3 means that the number of opinions in the population may increase by one. We note that no net increase occurs in step 3 if the new opinion is introduced by an individual whose previous opinion is not held by anyone else. In that case step 3 entails both an extinction and the introduction of a new opinion, which effectively means no change in the system other than a relabelling of one opinion. Thus, the number of opinions in the population after one combined iteration of steps 2 and 3 may be as before, $M(\tau+1)=M(\tau)$ if (i) no opinion goes extinct in step 2 and no new opinion is introduced in step 3, or (ii) if an opinion goes extinct (either in step 2 or in step 3, but not both), and a new opinion is introduced. Alternatively, we may have $M(\tau+1)=M(\tau)\pm 1$, if either an opinion goes extinct but no new one is introduced, or vice versa. 

\subsection{Definition of time units}

In the above we have used the notation $\tau$ to describe the number of microscopic (attempted) update steps that have occurred. In the following we will measure time in units of Monte Carlo steps (or generations). This time will be denoted by $t$, that is to say one unit of time $t$ corresponds to $N$ microscopic steps, or in other words, $t=\tau/N$. Unless specified otherwise any reference to time from now on will be with respect to time $t$, and similarly when we speak of a rate we will always mean the number of events per generation, that is per one unit of time $t$.

\subsection{Absence of absorbing states}
Due to possibility that new opinions emerge (for non-zero values of $\alpha$), the system cannot reach a permanent consensus state. This distinguishes the model from the traditional VM, which is recovered only for $\alpha=0$. Instead, for $\alpha>0$ the model with varying numbers of opinion remains in a dynamically active state indefinitely. This state can be characterised by the distribution of the number of opinions $M$ in the population, and the number of individuals $n_a$ in each opinion state ($a=1,2,\dots$). The variable $M$ can take values from $M=1$ (consensus) to $M=N$ (each agent is in a different opinion state). 

We also note that the present model is not equivalent to a noisy voter model \cite{carro2016noisy, peralta2018analytical, herrerias2019consensus} with $N$ possible opinion states. In the multistate noisy VM an agent can change opinion with some constant probability per iteration. Crucially, in this process the agent can adopt {\em any} opinion state, no matter whether any other agent holds that opinion or not. In step 3 of our model, the agent can only switch to an opinion that is not currently present in the population.
\begin{figure}
\includegraphics[width=1\linewidth]{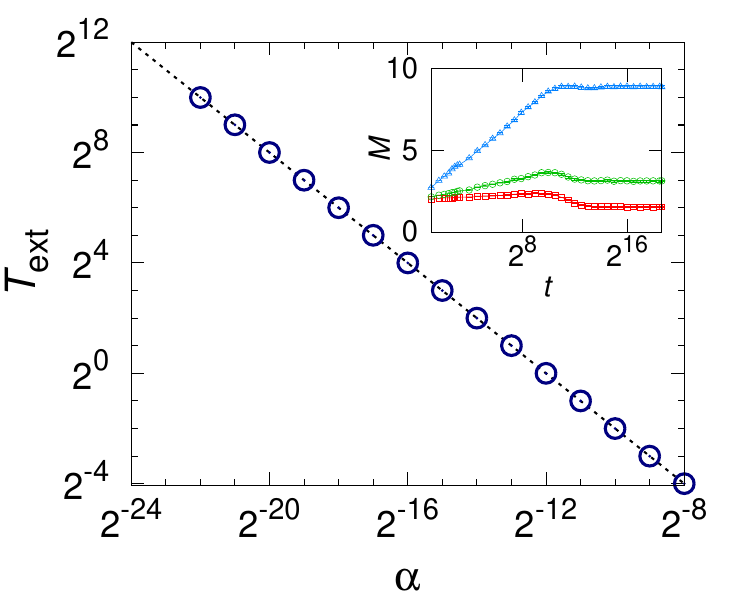}
\caption{Mean time $T_{\rm ext}$ between consecutive extinctions as a function of the innovation rate $\alpha$. Markers are from simulations for a population of size $N=2^{12}$ averaged over $10$ runs and the dashed line shows Eq.~(\ref{eq:T_ext}). 
(inset) Time evolution of the number of opinions $M$ in the population for $\alpha = 2^{-16}$ (red), $2^{-14}$ (green), and $2^{-12}$ (blue) averaged over $500$ runs. 
}
\label{fig:T_ext}
\end{figure}

\begin{figure*}
\includegraphics[width=\textwidth]{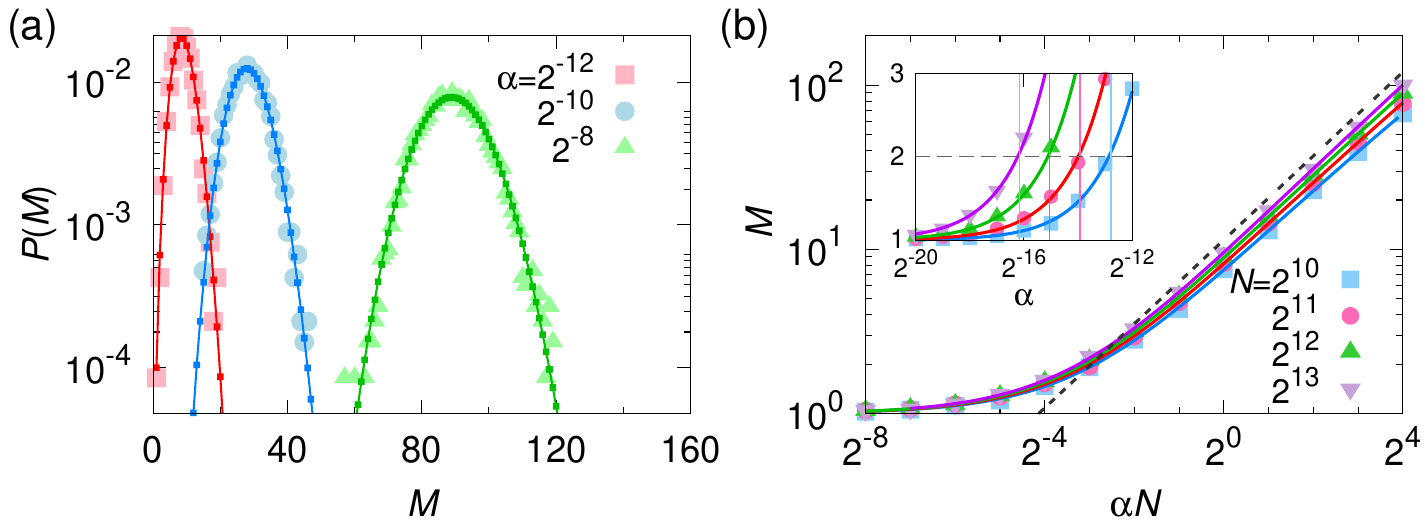}
\caption{
(a) The distribution $P(M)$ of the number of opinions in the steady state is shown for different values of $\alpha$. Numerical results (symbols) for $N=2^{12}$, averaged over $5000$ samples and the theoretical predictions (lines) from Eq.~(\ref{eq:pm}) are shown together. (b) Markers show the average number of opinions, $M$ as a function of $\alpha N$ for various population sizes $N$ as indicated. Lines show the theoretical predictions obtained by numerically evaluating the first moment of the distribution in Eq.~(\ref{eq:pm}). The inset shows a magnification for small $\alpha$ with various $N$ with the same symbols. The vertical lines in the inset represent the location of $\alpha N = 1/\log N$. The dashed line in panel (b) indicates power-law increase with an exponent of $0.85$ (obtained from a fit).
}
\label{fig:pm}
\end{figure*}

\section{Dynamics on Complete Graphs}

We begin by analyzing the VM with evolving opinions on a complete graph. This setup allows us to proceed analytically, at least in part, providing additional insight. 

\subsection{Number of opinions in the population}

\subsubsection{Balance of extinction and innovation}

Simulations show that the typical number $M$ of opinions in the population becomes constant after a transient period (see the inset of Fig.~\ref{fig:T_ext}). This means that the extinction rate (the expected number of extinctions per unit time) must balance the rate with which new opinions are introduced. Neglecting cases in which the introduction of a new opinion occurs simultaneously to the extinction of another in step 3 of the dynamics, the rate with which new opinions are introduced is $\alpha N$ (per unit time $t$). We thus expect the time interval $T_{\rm ext}$ between two consecutive extinctions of opinions to fulfill
\be\label{eq:T_ext}
T_{\rm ext}=\frac{1}{\alpha N}.
\ee
This relation is confirmed in numerical simulations in Fig.~\ref{fig:T_ext}. In addition, the average number of opinions, $M$, reaches a constant value as shown in the inset of Fig.~\ref{fig:T_ext}. In the following analysis, we focus on the properties of this stationary state.

\subsubsection{Distribution of the number of opinions}\label{sec:distrib_op}

Using an analogy to models of the evolution of mating types in \cite{berrios2021switching} we can calculate the distribution $P(M)$ of the number of different opinions present in the population in the stationary state. The imitation mechanism in the VM is akin to asexual reproduction in the evolutionary model in \cite{berrios2021switching}. In an asexual reproduction event an individual reproduces and generates an offspring which is of the same type as the parent. To keep the population size constant, a random other individual is removed from the population in \cite{berrios2021switching} following a reproduction. If this removed individual is of a type different to that of the reproducing individual such a combined reproduction-removal event has the same effect as an imitation event in our model -- an agent of one type (opinion) is replaced by an agent of another type. The rate for such events is indicated in Table I in \cite{berrios2021switching}, and is in direct analogy to step 2 of the update algorithm in Sec.~\ref{sec:setup} of the current paper. Similarly, the setup in \cite{berrios2021switching} allows for mutation, which is modelled as spontaneous state change of an individual to a mating type that is assumed not to be present in the population. This is analogous to spontaneous change of an agent to a new opinion not currently represented by any other agent. For further details see in particular Fig.~1 in \cite{berrios2021switching}, and the text in Sec. 2.1 of that paper.

The rate $\alpha N$ in our model is the rate with which new opinions appear. It plays the same role as the mutation rate $m_g$ in \cite{berrios2021switching}, the typical number of new mutations per generation. The number of opinions in our model is akin to the number of mating types.

Using this analogy, we then have, using Eq.~(15) in \cite{berrios2021switching},
\begin{align}\label{eq:pm}
P(M) = \frac{(N\alpha)^{M-1}}{(N-1)!} \frac{\scalebox{0.7}{$\begin{bmatrix} N \\ M \end{bmatrix}$}}{\scalebox{0.7}{$\begin{pmatrix} N\alpha+N-1 \\ N\alpha \end{pmatrix}$} },
\end{align}
where \scalebox{0.7}{$\begin{bmatrix} N \\ M \end{bmatrix}$} is the unsigned Stirling number of the first kind. This prediction is successfully tested against simulations for various values of $\alpha$  in Fig.~\ref{fig:pm}(a). From this distribution we can directly calculate the mean number of opinions in the population in the stationary state.
Although the binomial coefficient in Eq.~(\ref{eq:pm}) is formally defined only for integer arguments, we extend the equation to non-integer values of $N \alpha$ via the gamma function representation of factorials. In this way our result can be evaluated for non-integer values of $N \alpha$ .

\subsubsection{Typical number of opinions in the population}\label{sec:num_op}

The average number of opinions $M$ in the steady state is shown in Fig.~\ref{fig:pm}(b) as a function of $\alpha N$ for different population sizes. The number of distinct opinions grows with the rate $\alpha N$. The model exhibits two distinct behaviors: for small values of $\alpha$, the system is near consensus ($M\approx 1$). In this regime ($\alpha N \ll 1$), the introduction of new opinions is rare and if any new opinion emerges, it typically goes extinct before a further new opinion can appear. Conversely, for $\alpha N \gg 1$, the number $M$ of opinions increases approximately as a power law with increasing $\alpha N$.

We can estimate the range of innovation rates for which the system is near consensus as follows.  Assume that the system is in the near-consensus state, and that there is only one single minority opinion, held by a fraction $r_0$ of agents. Starting from this state, the consensus time for the conventional two-state VM is given by \cite{sood2008voter,vazquez2008analytical}
\be\label{eq:t_c_r0}
t_c(r_0) = N \left[ (1-r_0) \log \frac{1}{1-r_0}+r_0 \log \frac{1}{r_0} \right].
\ee
If the minority opinion is held only by one single agent, $r_0=1/N$, the consensus time is approximately $\log N$ for large $N$. 

 Thus, if the system has just left consensus, and is in a state with $M=2$ and the minority opinion is only held by one agent, the population will return to consensus approximately after $\log N$ units of time. Conversely, the rate with which new opinions appear is $\alpha N$, and if this happens the system will transition to $M=3$.  Thus, we can broadly expect that the point at which $\alpha N = 1/\log N$ corresponds to an average number of $M\approx 2$ opinions in the population. To test this estimate in simulations, we show the average number of opinions in the population for small $\alpha$ in the inset of Fig.~\ref{fig:pm}(b). The vertical lines represent the points at which $\alpha N = 1/\log N$. For innovation rates below this threshold the average number of opinions is indeed found to be below two, and for larger innovation rates we find $M\gtrsim 2$. We will return to this below in Sec.~\ref{sec:cons_frac}.

\subsection{Number of agents holding a particular opinion}\label{sec:num_agents}

\begin{figure*}
\includegraphics[width=\textwidth]{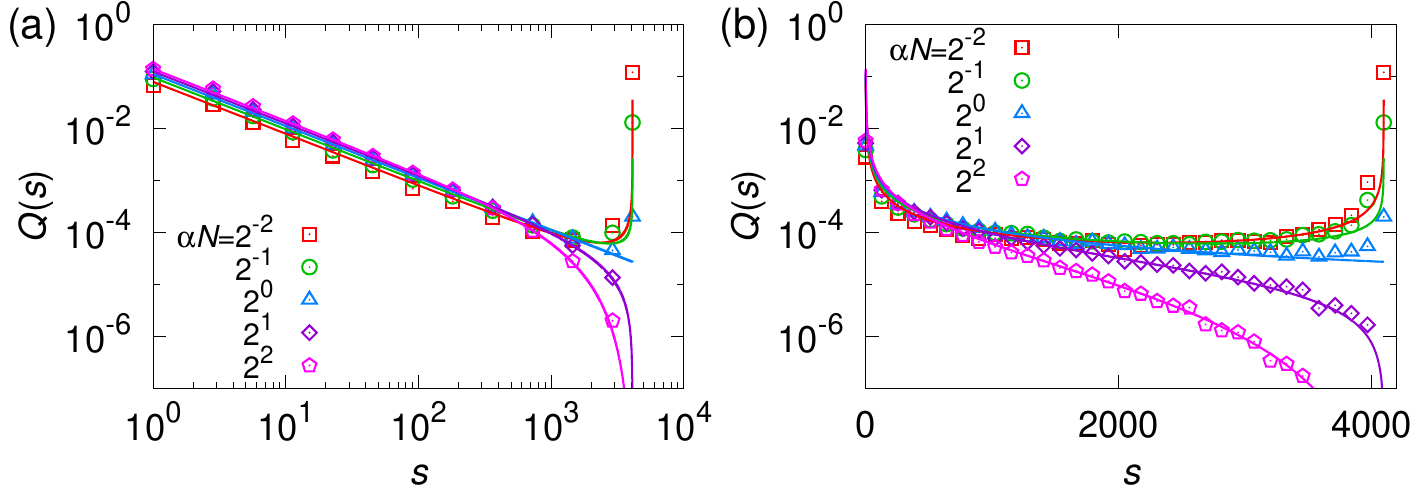}
\caption{The distributions $Q(s)$ of opinion sizes on complete graphs with $N=2^{12}$
are shown for different values of $\alpha N$ on a doubly logarithmic scale in panel (a), and in a linear-log representation in panel (b). The distributions decay with a power-law tail, following $Q(s)\sim s^{-1}$. Markers are simulation results, averaged over $5000$ realisations and lines represent the theoretical predictions obtained from Eqs.~(\ref{eq:qtilde_of_s}) and (\ref{eq:q_of_s}).}  
\label{fig:qs}
\end{figure*}

\subsubsection{Distribution of number of agents holding a particular opinion}\label{sec:Q_of_s}
We next study the distribution of the number of agents holding a particular opinion. In simulations this is measured as follows.  In the stationary state, we look at the system at a particular point in time. Then, we go through all opinions that are present at that point, and measure the number, $s_a$, of agents holding each of these opinions. By repeating this process for different realisations we obtain the distribution $Q(s)$.

Results are shown in Fig.~\ref{fig:qs}. A key feature of these distributions is their power-law decay at low values of $s$, specifically
following the form $Q(s) \sim s^{-1}$ as shown in Fig.~\ref{fig:qs}(a). This indicates that many opinions are represented by small number of agents. This is likely to be a consequence of the emergence of new opinions followed by the potentially relatively quick elimination of such a new opinion.

We also find that $Q(s)$ can either have a unimodal shape (for large $\alpha N$), or be bimodal (for small $\alpha N$).  Figure \ref{fig:qs}(b) shows a clear distinction between these outcomes. Unimodal distributions are found when new opinions are introduced frequently. The distribution $Q(s)$ exhibits a single peak at $s=1$, and then a power-law decay. Conversely, when innovation is rare two distinct peaks appear at $s=1$ and $s=N$, indicating that states near consensus occur frequently, in-line with the results in Fig.~\ref{fig:pm}(b).

We will now proceed to obtain an analytical approximation for $Q(s)$. To do this, we focus on a given opinion state that is present in the population, and formulate an effective one-step process for the number of agents holding that opinion. We follow the ideas of \cite{redner2019reality,herrerias2019consensus}. In order to track the number of agents in this state, we only need to distinguish between these agents and the agents who are in any other state. That is to say, if an agent is not in the focal state, then we do not need to know in what state exactly this individual is. The possibility to proceed in this way is a consequence of the linearity of the update rules.

The number of agents $s$ holding a particular opinion can be described by the following effective transition rates, 
\BE\label{eq:tpm_eff}
T^+(s)&=&\alpha N \delta_{s,0}+\frac{s(N-s)}{N}, \nonumber \\
T^-(s)&=&\alpha s + \frac{s(N-s)}{N}.
\EE
Here the rate $T^+(s)$ is for transitions $s\to s+1$, and $T^-(s)$ is for $s\to s-1$. These rates define a one-step process on the set $s\in\{0,1,2,\dots,N\}$. 
The first term in $T^+(s)$ indicates that $s$ changes from $0$ to $1$ with rate $\alpha N$. 
This is the `invention' of a previously non-existing opinion. Each agent does that with rate $\alpha$, so the overall rate is $\alpha N$. 
The second terms in $T^\pm(s)$ are the same as in the the normal VM dynamics. 
The first term in $T^-(s)$ (the term $\alpha s$) indicates that each of the $s$ agents holding 
a particular opinion can spontaneously invent a new opinion at rate $\alpha$ (and therefore leaves the focal state).

The stationary distribution $\tilde Q(s)$ of this process ($s=0, 1, \dots,N$) can be obtained in closed form
by standard methods (see e.g. \cite{ewens2004mathematical,herrerias2019consensus}). One has
\be\label{eq:qtilde_of_s}
\tilde Q(s)\equiv\frac{\prod_{k=1}^{s} \frac{T^+(k-1)}{T^-(k)}}
{1+\sum_{k=1}^{N} \prod_{\ell=1}^{k}\frac{T^+(\ell-1)}{T^-(\ell)}}.
\ee

This expression applies for $s=0,1,\dots,N$ (for $s=0$ the numerator takes the value of one). The denominator ensures normalisation. What we really measured in the simulations described in first paragraph of this Sec.~\ref{sec:Q_of_s} is the distribution of $s$ conditioned on the fact that the opinion is present in the system (i.e., that $s>0$). This object is, for $s>0$,
\be\label{eq:q_of_s}
Q(s)=\frac{\tilde Q(s)}{1- \tilde Q(s=0)}.
\ee
As seen in Fig.~\ref{fig:qs} this agrees well with results from simulations.

We now use these results to estimate where in parameter space the shape of the distribution $Q(s)$ transitions from unimodal to bimodal. For a given population size $N$, the value of $\alpha$ at which the distribution $Q(s)$ has zero slope at the right end in Fig.~\ref{fig:qs}, i.e., near $s = N$, can be obtained by setting $Q(s = N-1) = Q(s=N)$. Using stationarity, this translates into $T^{+}(N-1)=T^{-}(N)$. Using Eq.~(\ref{eq:tpm_eff}) this in turn means
\be
\alpha N = 1 - \frac{1}{N}.
\label{eq:an1}
\ee
For sufficiently large $N$, therefore, we expect the distribution $Q(s)$ to change from bimodal to unimodal shapes roughly when $\alpha N=1$. This is consistent with the data shown in Fig.~\ref{fig:qs}. Broadly speaking, the change of shape of $Q(s)$ at $\alpha N\approx 1$ indicates that consensus is unlikely to occur for $\alpha N>1$, but that consensus can be observed with non-zero probability for $\alpha N <1$.

\subsubsection{Fraction of agents with the majority opinion}

\begin{figure}
\includegraphics[width=\linewidth]{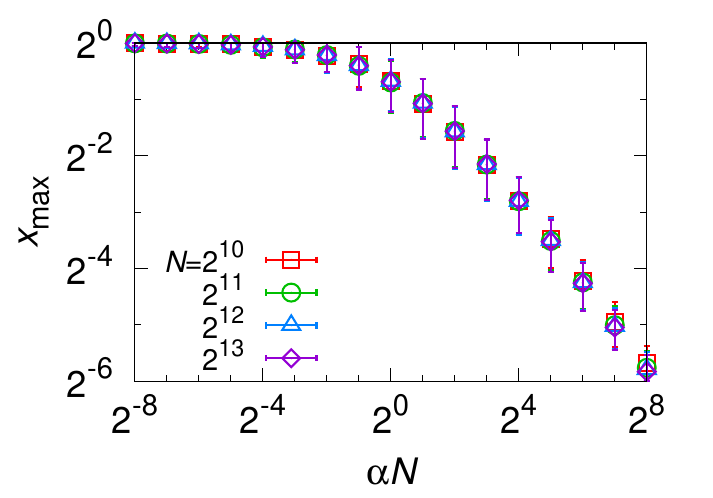}
\caption{Typical fraction of agents holding the majority opinion, $x_{\max}$, as a function of $\alpha N$. 
Markers are from simulations of populations with all-to-all interaction with different population 
sizes. 
\label{fig:max_x}
}
\end{figure}

We have further measured  the mean fraction of agents 
holding the majority opinion, that is
\be
x_{\max}\equiv \mbox{max}_a \frac{n_a}{N},
\ee
where $n_a$ is the number of agents holding opinion $a$.
Results are shown in Fig.~\ref{fig:max_x}. This quantity is close to one for $\alpha N\ll 1$, i.e., nearly all agents in 
the population hold the same opinion. This behavior is consistent with 
the bimodal distributions of $Q(s)$ in Fig.~\ref{fig:qs}(b), where one of the peaks 
occurs at $s = N$. 
It also agrees with the result in Fig.~\ref{fig:pm}(b), where $M \lesssim 2$ when $\alpha N < 1/\log N$.  
For $\alpha N\gg 1$, however, $x_{\rm max}$ is found 
to be much smaller than one, and is a decreasing function of $\alpha N$. 
This is consistent with a picture in which each opinion is only 
held by a small number of agents.
It also reflects the unimodal distributions in Fig.~\ref{fig:qs},
showing a single peak of $Q(s)$ at $s=1$.

\subsection{Fraction of time spent in the consensus state}\label{sec:cons_frac}

Following on from Secs.~\ref{sec:num_op} and \ref{sec:num_agents} we study the fraction of time the system remains in consensus as a function of $\alpha N$. Specifically, we measured the total time spent in any of the consensus states, relative to the total simulation time (in the stationary state), i.e., $t_{\rm cons}/t_{\rm tot}$. 
This quantity is the probability to find the system in consensus at any random time in the stationary state, given by $P(M=1)$ in Eq.~(\ref{eq:pm}). Using the identity $\begin{bmatrix} N \\ 1 \end{bmatrix}=(N-1)!$ we find
\begin{align}
P(M=1) = \frac{1}{\binom{N \alpha +N-1}{N \alpha}}.
\end{align}
By examining this fraction for different values of $\alpha N$, we can identify the conditions under when a consensus becomes possible with non-vanishing probability, and when it becomes the prevailing outcome.

\begin{figure}
\includegraphics[width=\linewidth]{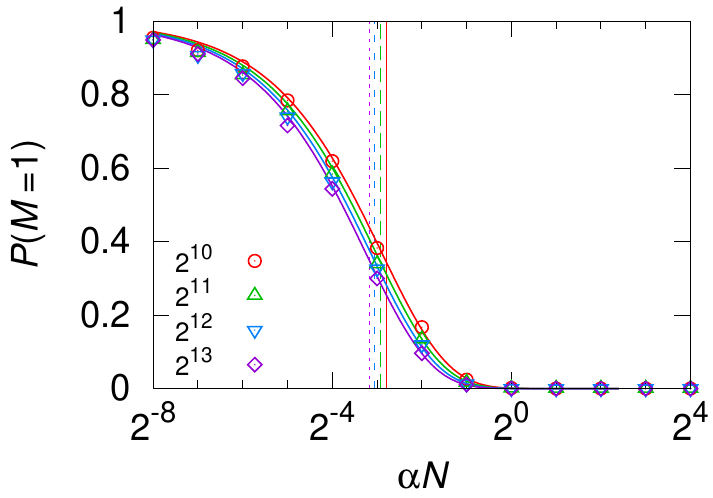}
\caption{Probability to find the system in consensus in the stationary state, that is measured as $t_{\rm cons}/t_{\rm tot}=P(M=1)$, where $t_{\rm cons}$ is the total time (in the stationary state) spent in consensus, and $t_{\rm tot}$ is the total simulated time in the stationary state. Markers are from simulations and lines are from the theoretical predictions in $P(M=1)$ in Eq.~(\ref{eq:pm}). The vertical lines indicate the point $\alpha N= 1/\log N$ for different $N$. 
\label{fig:tcons}
}
\end{figure}

Results are shown in Fig.~\ref{fig:tcons}. When $\alpha N>1$ we find $t_{\rm cons}/t_{\rm tot}$ is zero to all intents and purposes, that is the system is never found in consensus. However, as $\alpha N$ decreases below unity, the fraction of time spent in consensus becomes non-zero and gradually increases. This indicates that consensus can appear with non-zero probability in line with the result in Eq.~(\ref{eq:an1}). When $\alpha N$ decreases further below $1/\log N$, consensus states gradually persists for longer. We note that the point $\alpha N=1/\log N$ moves to the left on the $\alpha N$ axis as $N$ increases. From this we expect that, at fixed $\alpha N<1$, the fraction of time spent in consensus would decrease with increasing population size $N$. This is because as the system grows, the time to reach consensus increases and is consistent with the numerical results in Fig.~\ref{fig:tcons}.

\subsection{Density of active interfaces}

We next calculate the time evolution and asymptotic value of the density of active links in the system
(we focus on all-to-all connectivity, thus the word link refers to connections between any two nodes).
Our calculation assumes an infinite system size, but other than that no approximation is made. The density of active links is the proportion of links connecting agents who are in different opinion states. We write $\rho(t)$ for this density at time $t$.

The number of active links can change via imitation events, and it increases when a new opinion appears in the population (unless there is a simultaneous extinction). If an individual of type $a$ imitates opinion $b$ the number of interfaces in a population with all-to-all interaction changes by $(n_a-1)-n_b$, where $n_a$ and $n_b$ are the numbers of individuals in the population with opinions $a$ and $b$, respectively, before the event. Such an event occurs with rate $n_a n_b / (N-1)$ (where time is measured in generations). If, on the other hand an individual who is currently of type $a$ introduces a new opinion, the number of interfaces increases by $n_a-1$. This event occurs with rate $\alpha n_a$.

\begin{figure}
\includegraphics[width=\columnwidth]{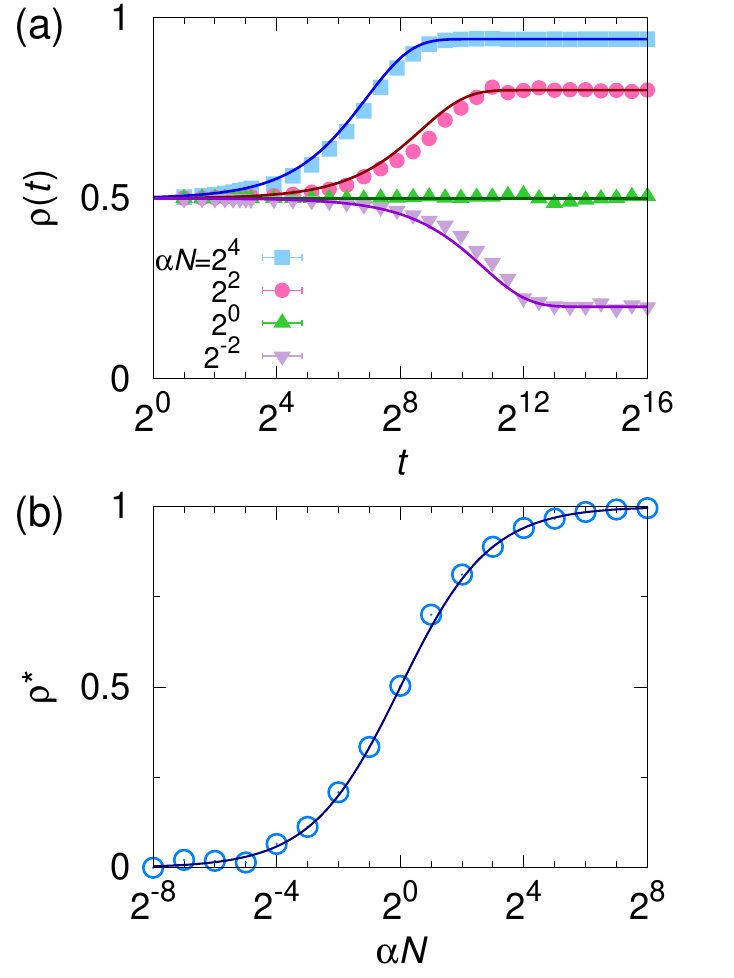}
\caption{
(a) Time evolution of the active link density $\rho(t)$ for different values of $\alpha N$, and 
(b) asymptotic density $\rho^*$ of active link in the steady state with $N=2^{12}$ as a function of $\alpha N$.
Markers are from numerical simulations with $N=2^{12}$ and lines are from Eq.~(\ref{eq:rho_analytical}) and the expression for $\rho^*$ given in the text.
}
\label{fig:rho}
\end{figure}

Putting things together, and writing $L=N(N-1)/2$ for the total number of links in the well-mixed system, 
\begin{align}
\frac{d\rho}{dt} &= \frac{1}{L} \left[ \sum_{a<b} \left\{ \frac{n_a n_b}{N-1}[(n_a-n_b-1)+(n_b-n_a-1]) \right\} \right.\nonumber \\
&\left.+ \alpha \sum_a n_a (n_a-1) \right]. \nonumber \\
&=\frac{1}{L} \left[ \sum_{a<b}  \frac{-2n_a n_b}{N-1}+ \alpha \sum_a n_a (n_a-1) \right].
\end{align}
The notation $\sum_{a<b}\dots$ indicates a sum over $a$ and $b$, covering all pairs $a,b\in\{1,\dots,M\} $ such that $a<b$. Using the relations
\begin{align}
\rho = \frac{ \sum_{a<b} n_a n_b}{L}, \quad
1-\rho = \frac{ \sum_{a} [n_a (n_a-1)/2]}{L},
\end{align}
we finally arrive at the following differential equation
\begin{align}
\frac{d\rho}{dt} &= -\frac{2}{N-1} \rho + 2 \alpha (1-\rho).
\label{eq:rate}
\end{align}
The solution of Eq.~(\ref{eq:rate}) is given by
\begin{align}\label{eq:rho_analytical}
\rho(t) = e^{-t/\delta} (\rho_0 - \rho^*) + \rho^*,
\end{align}
with $\rho^*= \frac{\alpha(N - 1)}{1 + \alpha(N - 1)}$ and $\delta = \rho^*/(2\alpha)$.
Thus, $\rho$ decays towards the asymptotic value $\rho^*$, with decay time $\delta$. The density $\rho^*$ of active links in the steady state is non-zero unless $\alpha=0$ (in this latter case the model
becomes the identical to the conventional VM). From the expression for $\rho^*$ we note that $\rho^*=1/2$ when $\alpha(N-1)=1$. Fig.~\ref{fig:rho}(b) indicates a cross-over of the stationary density of active interfaces from $\rho^*=0$ for small $\alpha N$ to $\rho^*=1$ for large $\alpha N$. Thus, $\rho^*$ takes the separating value $\rho^*=1/2$ when $\alpha (N-1)=1$. For large $N$ this reduces to $\alpha N=1$, and hence to the condition at which the quasi-stationary distribution $Q(s)$ changes shape [Eq.~(\ref{eq:an1})]. The same condition also comes up in Sec.~\ref{sec:cons_frac}.

In Fig.~\ref{fig:rho}, we present simulation results to confirm the validity of our calculation.  Panel (a) shows the time-evolution of $\rho(t)$ for different choices of $\alpha N$, and in panel (b) we plot the asymptotic density $\rho^*$ of active links. 
We confirm that the results from the rate equations are in good agreement with simulations, both for the time evolution $\rho(t)$ and the interface density $\rho^*$ in the steady state.

The asymptotic density of active interfaces is close to 
one for $\alpha N \gg 1$, when new opinions are introduced at a very high rate. To a good approximation, each agent holds their own opinion. All interfaces are then active, and we find $\rho\approx 1$. Indeed, one can find that $\rho^* \approx 1- 1/[\alpha(N-1)]$ for $\alpha \gg 1/(N-1)$. For $\alpha N \ll 1$ on the other hand, the introduction of new opinions is rare, relative to the timescales of the imitation dynamics. This means that the system has sufficient time to organize itself near consensus between innovation events. Consequently, we find that the density of active interfaces is close to zero. From the analytical solution for $\rho^*$ we have $\rho^* \approx \alpha(N-1)$ when $\alpha \ll 1/(N-1)$.

\section{Dynamics on random networks}

\begin{figure}
\includegraphics[width=\columnwidth]{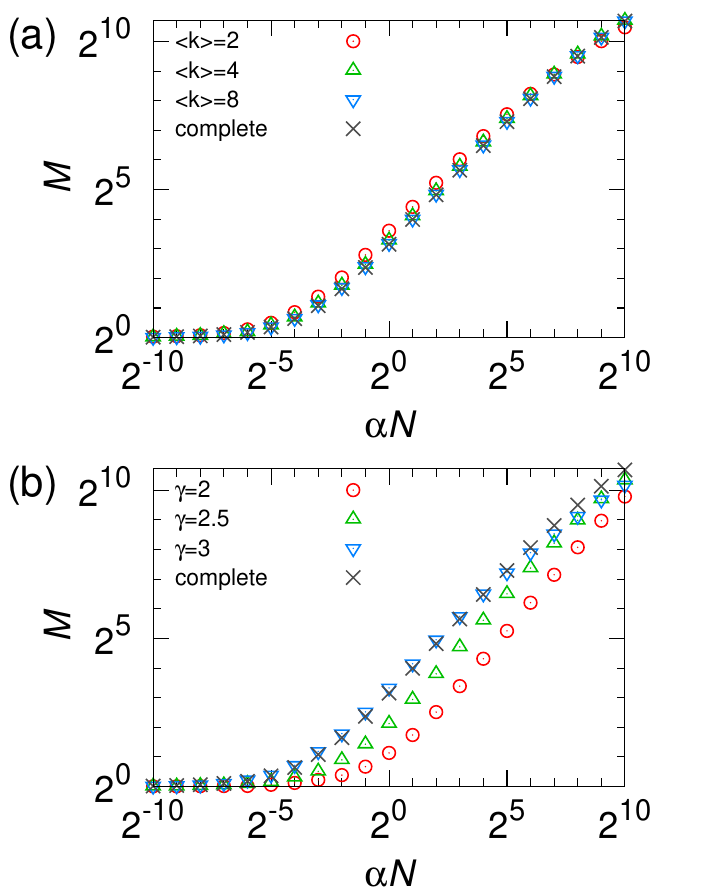}
\caption{Typical number of opinions $M$ as a function of $\alpha N$. Panel (a) is for Erd\H{o}s-R\'enyi networks with mean degrees $\langle k \rangle=2,4,8$ and panel (b) is for scale-free networks with degree exponents $\gamma=2,2.5,3$. The markers are from simulations on graphs of size $N=2^{12}$. Results for complete graphs are shown for comparison.
}
\label{fig:net}
\end{figure}

We conducted Monte Carlo simulations of the model on random networks consisting of $N$ nodes to study
how network topology influences the formation of opinions. We examined the number of opinions 
$M$ as a function of $\alpha N$ in two types of networks: Erd\H{o}s-R\'enyi (ER) networks 
with varying mean degrees $k=2,4,8$ [Fig.~\ref{fig:net}(a)] and scale-free (SF) networks constructed using 
configuration models with degree exponents $\gamma=2,2.5,3$ [Fig.~\ref{fig:net}(b)]. This setup allowed
us to analyze the impact of network density and degree heterogeneity.

The behavior of $M$ in ER networks is similar to that for complete graphs, regardless of network density, 
as shown in Fig.~\ref{fig:net}(a). In contrast, SF networks display distinct patterns depending on
the network heterogeneity shaped by the degree exponent $\gamma$. Specifically, as $\gamma$ 
decreases, indicating higher heterogeneity in degree distributions, the number of opinions $M$ decreases. 
This suggests that structural heterogeneity plays a role in determining how opinions 
form and evolve within populations.

While the innovation process is independent of the structure of the interaction network, the consensus time $t_c$ 
is mainly determined by the moments of the degree distribution, following the 
relationship, $t_c \propto N \langle k \rangle^2 / \langle k^2 \rangle$ \cite{sood2005voter,sood2008voter,vazquez2008analytical}. This indicates that degree heterogeneity shortens 
the time to consensus (i.e., consensus is formed more rapidly on heterogeneous networks). At fixed innovation rate greater degree heterogeneity thus leads to fewer opinions being present in the population.  This occurs because hubs in the network dominate the dynamics, suppressing the diversity of opinions by bringing the network to consensus or near-consensus states.

\section{Discussion}

To summarize, we have introduced and studied a modified voter model with an evolving number of opinions. New opinions can emerge and existing ones may fade over time. Unlike in the conventional voter model, the dynamics have no absorbing states for non-zero innovation rate. We found that the balance between innovation and extinction shapes the number of opinions in the steady state. Using a combination of analytical approximations and numerical simulations, we identified distinct behaviors of the models depending on innovation rate $\alpha$ and the size of the population, $N$. 
For $\alpha N  \lesssim 1/\log N$, the system is very frequently in one of the consensus states. For $1/\log N \lesssim \alpha N \lesssim 1 $ consensus is seen with non-zero probability in the stationary state, but it is not the dominant outcome. For $\alpha N \gtrsim 1$ innovation is so quick that one practically never finds consensus. In-line with this, the distribution $Q(s)$ is bimodal (peaks at $s=1$ and $s=N$) for $\alpha N<1$, and unimodal (single peak at $s=1$) for $\alpha N>1$. We also found that network structure affects the outcome. In particular degree heterogeneity leads to quicker consensus formation, and hence reduces the number of opinions that are present in the system.

The proposed model shares similarities with the noisy voter model \cite{carro2016noisy,peralta2018analytical,herrerias2019consensus}. However, there are also differences in that an agent in the noisy VM can spontaneously change to one of any possible opinion, whereas in our model spontaneous changes are only possible to opinions that are not currently present in the population. Parallels can also be drawn with models of biological evolution, and in particular those of the evolution of mating types. \cite{Constable,berrios2021switching}. In these models birth, death, and mutation processes govern diversity of species and/or mating types. Mathematically our model is very similar. These connections are an example of the parallels between models of social and biological dynamics, highlighting that it is possible, as we have done, to use analytical methods developed in one area in the respective other field.

Our model could be developed further, for example to include heterogeneous innovation rates for different agents, adaptive networks or additional sources of noise such as varying external conditions or periods with high and low innovation rates respectively. Work along these lines could contribute to our understanding of the mechanisms driving consensus and diversity in the dynamics of complex individual-based systems.

\begin{acknowledgments}
This research was supported in part by the National Research Foundation of Korea (NRF) grant funded by the Korea government (MSIT) (No. 2020R1I1A3068803) and by Global - Learning \& Academic research institution for Master’s $\cdot$ PhD students, and Postdocs (LAMP) Program of the National Research Foundation of Korea (NRF) grant funded by the Ministry of Education (No. RS-2024-00445180).
This work was also supported by a funding for the academic research program of Chungbuk National University in 2024.  Partial financial support has been received from Grant PID2021-122256NB-C21, PID2021-122256NB-C22 funded by MICIU/AEI/ 10.13039/501100011033 and by “ERDF/EU”, and the Mar{\'i}a de Maeztu Program for units of Excellence in R\&D, grant CEX2021-001164-M. BM and TG thank Maxi San Miguel for bringing us together in a meeting, where this work started.

\end{acknowledgments}

\bibliography{voter_evolving}

\end{document}